\documentclass[superscriptaddress,aps, prb, twocolumn,showpacs,showkeys,notitlepage,amsmath, amssymb]{revtex4-2}

\usepackage[utf8]{inputenc}
\usepackage[varg]{txfonts}

\usepackage{graphicx}
\usepackage[caption=false]{subfig}
\usepackage[mathlines]{lineno}
\usepackage[svgnames]{xcolor}
\usepackage{hyperref}
\usepackage{color}
\usepackage{rotating}
\usepackage{comment}
\usepackage{threeparttable}
\usepackage[referable]{threeparttablex}
\usepackage{siunitx}
\usepackage{pbox}
\usepackage{braket}
\usepackage{physics}
\usepackage{nicefrac}
\usepackage{amsmath}

\usepackage{booktabs,microtype,afterpage} 
\hypersetup{
	colorlinks = true,
	linkcolor = Blue,
	citecolor = Blue,
	urlcolor  = Blue
}
\usepackage{amsmath}

\renewcommand{\vec}[1]{\mathbf{#1}}
\newcommand{\na}{Na\textsubscript{2}PrO\textsubscript{3}}

\def \na {Na$_{2}$PrO$_{3}$}
\def \muon {$\mu$SR}

\begin{document}
\title{Unraveling the magnetic ground-state in alkali-metal lanthanide oxide \texorpdfstring{\na}{}}

\author{\hspace{1mm}Ifeanyi John Onuorah}

\affiliation{
Dipartimento di Scienze Matematiche, Fisiche e Informatiche, Universit\'a di Parma, I-43124 Parma, Italy
}

\author{\hspace{1mm}Jonathan Frassineti}

 \affiliation{
Dipartimento di Fisica e Astronomia "A. Righi", Universit\'a di Bologna, I-40127 Bologna, Italy }

\author{\hspace{1mm}Qiaochu Wang}
\affiliation{
Department of Physics, Brown University, Providence, Rhode Island 02912, USA}

\author{\hspace{1mm}Muhammad Maikudi Isah}
\affiliation{
Dipartimento di Fisica e Astronomia "A. Righi", Universit\'a di Bologna, I-40127 Bologna, Italy }

\author{\hspace{1mm}Pietro Bonf\`a}
\affiliation{
Dipartimento di Scienze Matematiche, Fisiche e Informatiche, Universit\'a di Parma, I-43124 Parma, Italy
}

\author{\hspace{1mm}Jeffrey G. Rau}
\affiliation{
Department of Physics, University of Windsor, Windsor, Ontario, Canada N9B 3P4
}

\author{J. A. Rodriguez-Rivera}
\affiliation{NIST Center for Neutron Research, National Institute of Standards 
    and Technology, Gaithersburg, MD 20899, USA}
\affiliation{Department of Materials Science and Engineering, University of 
    Maryland, College Park, MD 20742, USA}       
\author{A. I. Kolesnikov}
\affiliation{Neutron Scattering Division, Oak Ridge National Laboratory, Oak Ridge, Tennessee 37831, USA}

\author{\hspace{1mm}Vesna F. Mitrovi{\'c}}
\affiliation{
Department of Physics, Brown University, Providence, Rhode Island 02912, USA}

\author{\hspace{1mm}Samuele Sanna}
 \affiliation{
Dipartimento di Fisica e Astronomia "A. Righi", Universit\'a di Bologna, I-40127 Bologna, Italy }

\author{\hspace{1mm}Kemp W. Plumb}
\affiliation{
Department of Physics, Brown University, Providence, Rhode Island 02912, USA}

\date{\today}

\begin{abstract}
  A comprehensive set of  muon spin spectroscopy and neutron scattering measurements supported by ab-initio and model Hamiltonian simulations have been used to investigate the magnetic ground state of \na{}. $\mu$SR reveals a N\'eel antiferromagnetic order below $T_N\! \sim\! 4.9$ K, with a small static magnetic moment  $m_{\rm static}\!\leq \! 0.22$~$\mu_{\rm B}/{\rm Pr}$ collinearly aligned along the $c-$axis. Inelastic neutron measurements reveal the full spectrum of crystal field excitations and confirm that the Pr$^{4+}$ ground state wave function  deviates significantly from the $\Gamma_7$ limit that is relevant to the Kitaev model. Single and two magnon excitations are observed in the ordered state below $T_N=4.6$~K and are well described by non-linear spin wave theory from the N\'eel state using a magnetic Hamiltonian with Heisenberg exchange $J=1$~meV and symmetric anisotropic exchange $\Gamma/J=0.1$, corresponding to an XY model. Intense two magnon excitations are accounted for by $g$-factor anisotropy $g_\mathrm{z}/g_\pm = 1.29$. A fluctuating moment $\delta m^2 = 0.57(22)$~$\mu_{\rm B}^2/{\rm Pr}$ extracted from the energy and momentum integrated inelastic neutron signal is reduced from expectations for a local $J=1/2$ moment with average $g$-factor $g_{\rm avg}\approx 1.1$. Together, the results demonstrate that the small moment in \na{} arises from  crystal field and covalency effects and the material does not exhibit significant quantum fluctuations. 
\end{abstract}

\maketitle

\section{Introduction}
Material realizations of Heisenberg-Kitaev models hold significant contemporary interest because of their  potential to harbor quantum spin liquids in higher than one-dimensional lattice geometries \cite{Jackeli2009, Takagi2019, Motome_2020}. The Kitaev model is  characterized by quantum frustration arising from bond-dependent anisotropic Kitaev interactions between $J_{\mathrm{eff}}=1/2$ Kramer's doublets spin-orbit entangled local magnetic moments \cite{Khaliullin2005, Jackeli2009}. In practice, the generic Hamiltonian for a realistic Kitaev material contains Heisenberg and symmetric anisotropic exchange interactions.  Depending on local symmetries and electron hopping pathways  these additional interactions  may be of comparable strength to the Kitaev term \cite{rau2014}. Dominant Kitaev interactions are known to be present in transition metal compounds with octahedrally coordinated 5$d^5$ and 4$d^5$ electronic filling, including the Ir$^{4+}$ oxides and  Ru$^{3+}$ based chlorides and trihalides \cite{ PhysRevB.94.041101,  PhysRevLett.105.027204, PhysRevLett.108.127203,  PhysRevB.92.235119, sears2020, PhysRevB.101.100410, Takagi2019, TREBST20221, PhysRevB.90.041112, Banerjee2017}, but strong Heisenberg and anisotropic interactions stabilize N\'eel order in all of these compounds \cite{rau2014}. It is thus essential to search for other material realizations where the non-Kitaev interaction terms can be further minimized.

Recently, it has been shown that strong Kitaev interactions are potentially realizable in a broader class of materials, including compounds with high spin $d^7$ electronic configuration such as Co$^{2+}$ and Ni$^{3+}$  \cite{PhysRevB.97.014407, PhysRevB.97.014408}, and the $f^1$ electronic configuration such as Ce$^{3+}$ and Pr$^{4+}$ \cite{ PhysRevB.99.241106, Motome_2020}. In the case of the $f^1$ electron materials,  a dominant spin-orbit coupling and octahedral crystal field can act to stabilize a $\Gamma_7$ doublet, $J_\mathrm{eff} = 1/2$, with strong Kitaev interactions \cite{Motome_2020, PhysRevMaterials.4.104420}. Furthermore, the more spatially localized  $f$-orbital occupying electrons compared to those in $4d$ and $5d$ orbitals could serve to limit direct exchange interactions.

One candidate material to realize such Kitaev physics is \na{}, which hosts octahedrally coordinated Pr$^{4+}$ in the $f^1$ electronic configuration \cite{HINATSU2006155}. \na{} crystallizes in the C$2$/c space group with edge-sharing PrO$_6$ octahedra, required for the realization of the dominant bond directional exchange \cite{PhysRevB.99.241106, PhysRevMaterials.4.104420, Motome_2020}.  The crystal structure has two inequivalent Pr sites forming a honeycomb lattice in the $ab$ plane with two intraplane Pr$-$Pr distances of $d\!=\!3.433$ and $d^{\prime}\!=\! 3.458$~\AA{}. Honeycomb planes are separated by layers of Na atoms with an interplane Pr$-$Pr distance of $d_{p}$ = 5.867 \AA{} \cite{ramanathan2021, PhysRevB.103.L121109}. Based on {\it ab initio} studies, \na{}  was predicted to host antiferromagnetic Kitaev interactions between local J$_{\mathrm{eff}}=1/2$ moments on Pr$^{4+}$ \cite{PhysRevB.99.241106, PhysRevMaterials.4.104420}. On the contrary, neutron crystal field measurements revealed that the Pr$^{4+}$ ground state wavefunction is not a $\Gamma_7$ doublet as required for the Kitaev model. Furthermore, the magnetic excitations are best captured by a Heisenberg XXZ model Hamiltonian with negligible Kitaev interactions \cite{PhysRevB.103.L121109}. Recent neutron and x-ray absorption measurements confirm that the Pr ground state in \na{} deviates significantly from the $\Gamma_7$ doublet as a result of strong octahedral crystal field and Pr(4$f$)$-$O(2$p$) hybridization \cite{PhysRevB.103.L121109, Ramanathan2023}.  However, there are still questions regarding the magnetic ground state of \na{}. Despite a clear heat capacity peak at $T_N$ \cite{HINATSU2006155}, and the appearance of well defined spin waves, no direct evidence for static magnetic order has been found in this material. Furthermore, an apparent continuum of magnetic excitations existing above the spin wave bands remains unexplained \cite{PhysRevB.103.L121109}. Together, the conspicuously small static magnetic moment and excitation continuum leave open the possibility for significant quantum fluctuations in \na{}. 

In this work, we use a comprehensive series of muon spin spectroscopy (\muon) and neutron scattering measurements accompanied by DFT and model Hamiltonian simulations to refine the magnetic ground state and underlying microscopic exchange interactions of \na{}. Our results rule out any significant, beyond spin-wave, quantum fluctuations in this material and establish crystal field effects and stacking faults as the primary origin of the small ordered moment. The \muon{} measurements unambiguously reveal that \na{} is antiferromagnetically ordered below $T_N\!\approx\!4.9$~K. Analysis of the muon data supplemented with DFT and dipolar simulations find  the most likely ordered state to be a  N\'eel AF structure with a small static magnetic moment of  $\mu_\mathrm{static}\! \leq \! 0.22$~$\mu_{\rm B}/Pr$,  collinearly aligned along the $c-$axis. Measurements of the crystal  field excitations reveal five additional crystal field modes above the previously reported single crystal field excitation \cite{PhysRevB.103.L121109}. The full set of crystal fields more tightly constrain the Pr$^{4+}$ ground state wavefunction. Low energy inelastic neutron scattering revealed intense single and multi-magnon excitations. The complete measured magnetic excitation spectrum is well described by a non-linear spin wave model in the collinear N\'eel state that includes a dominant Heisenberg exchange and subdominant  symmetric anisotropic  exchange interactions. We find that the previously unexplained continuum of magnetic excitations arises from a multi-magnon excitations with  relatively large neutron intensity  that is accounted for by $g$-factor anisotropy. Our analysis demonstrates a small ($\sim$20\%) ordered moment reduction arising from quantum fluctuations. The fluctuating moment of $\delta m = \sqrt{g^2J(J+1)- g^2J_z^2} = 0.75(14)$~$\mu_{\rm B}/{\rm Pr}$ recovered from the energy and momentum integrated inelastic neutron intensity is reduced from expectations for a local $J=1/2$ and $g_{\rm avg}=1.1$ as determined by our crystal field analysis.  Based on these results, we attribute the small moment and intense multi-magnon neutron intensity  in \na{} to crystal field and covalency effects.   

The remainder of this paper is organized as follows. After an overview of the methods in Sec.~\ref{sec:methods}, we present the \muon{} measurements in Sec.~\ref{sec:muon} along with  DFT and dipolar simulations of the muon spectra in Sec.~\ref{sec:mucalc}. In Sec.~\ref{sec:cf_neutron} we present and analyze the crystal field excitations.  Elastic and inelastic neutron scattering  are presented in Secs.~\ref{sec:elastic_neutron} and \ref{sec:spinwave_neutron} along with a non-linear spin wave modeling of the measured spectrum, that is used to constrain a model Hamiltonian. The summary and conclusions are presented in Sec.~\ref{sec:discussion}.

\section{Experimental and Computational Methods}\label{sec:methods}
Powder samples of \na{} were synthesized via solid-state reactions from Na$_2$O$_2$ and Pr$_6$O$_{11}$.  Dry starting reagents were weighed in a metal ratio, Na/Pr$\sim$2.2, to account for sodium evaporative losses, ground in an agate mortar and pestle, and pelletized under an argon environment. The prepared materials were enclosed in Ag ampules and heated at 750$^{\circ}$C under dry, flowing oxygen for 36 hours. Samples were furnace-cooled to $\sim\!150^{\circ}$C and immediately transferred to an Argonne glovebox for storage.

\muon{} measurements were carried out on the GPS spectrometer at the Paul Scherrer Institut, Switzerland. The sample was packed into aluminium foil inside a glove-box to avoid air contamination and put into a Cu fork inserted into the experimental probe. The measurements were performed both in a weak Transverse Field (TF) mode to calibrate the asymmetry parameter of the muon polarization, and in Zero Field (ZF) to reveal presence of the internal magnetic field that gives rise to the spontaneous oscillations of \muon{} signal. The ZF \muon{} spectra were collected at temperatures ranging from 1.5 K to 5.2 K using a helium flow cryostat. The time-differential \muon{} data were fitted using MUSRFIT software \cite{SUTER201269}, and the MuLab suite, a home-built Matlab toolbox.

To identify the muon implantation sites, we used the well-established DFT+$\mu$ approach \cite{Moller_2013, bonfa2016, onuorah2018, blundell2023}. Non-spin polarized DFT calculations were performed within generalized gradient approximation (GGA) for the PBE  (Perdew-Burke-Ernzerhof ~\cite{pbe1996}) exchange-correlation functional as implemented in the Quantum Espresso code~\cite{qe2009}. The muon was treated as a hydrogen impurity in a 2$\times$1$\times$1 charged supercell (96 host atoms and 1 muon) with a compensating background.  For Na, O, and H atomic species, the norm-conserving pseudopotentials were used;  while for Pr, the Projector augmented wave (PAW) with no 4$f$ electron state in the valence was used, in order to avoid the well-known difficulty in describing its valence shell \cite{PhysRevB.50.17953}. The plane-wave cut-off of 100 Ry  was used while the Brillouin zone was sampled using a 4$\times$4$\times$4 mesh of k-points~\cite{kpoint1976}.

High energy neutron scattering measurements of the crystal field excitations were conducted on the fine-resolution Fermi chopper spectrometer (SEQUOIA) \cite{Granroth:2010} at the Spallation Neutron Source (SNS), Oak Ridge National Lab  (ORNL). The polycrystalline sample was loaded into an  Aluminum can, and cooled to a base temperature of $T\!=\!5$~K using a closed cycle cryostat.  To cover the full range of crystal field exctiations with sufficient resolution, data was collected using the high-flux chopper with fixed incident neutron energies of $E_{\rm i}=$~60, 150, 300, 700, and 2500~meV.
 
Low energy neutron scattering measurements were carried out on powder samples using the Multi-Axis Crystal Spectrometer (MACS) at the NIST Center for Neutron Research (NCNR). Elastic ($E\!=\!0$) measurements were conducted with the monochromator in a vertical focusing configuration using neutron energy of 5~meV. Inelastic measurements were carried out using a double-focusing configuration and a fixed final energy of $E_{\rm f}$=3.7~meV with Be and BeO filters before and after the sample, respectively for energy transfers $\Delta E\!=\!E_{\rm i}\!-\!E_{\rm f}$ below 1.5~meV. For energy transfers above 1.4~meV, we used a Be filter after the sample and no incident beam filter. Data for energy transfers above 1.4~meV was corrected for contamination from high-order harmonics in the incident beam neutron monitor.  Measured background signal contributions from the sample environment were subtracted and signal count rates were converted to  absolute values of the scattering cross-section using the incoherent signal from the sample integrated over the range $1.7\!\leq\!Q\!\leq\!2$~\AA$^{-1}$. 

\begin{figure*} [htbp]
\includegraphics[width = 1.0\linewidth]{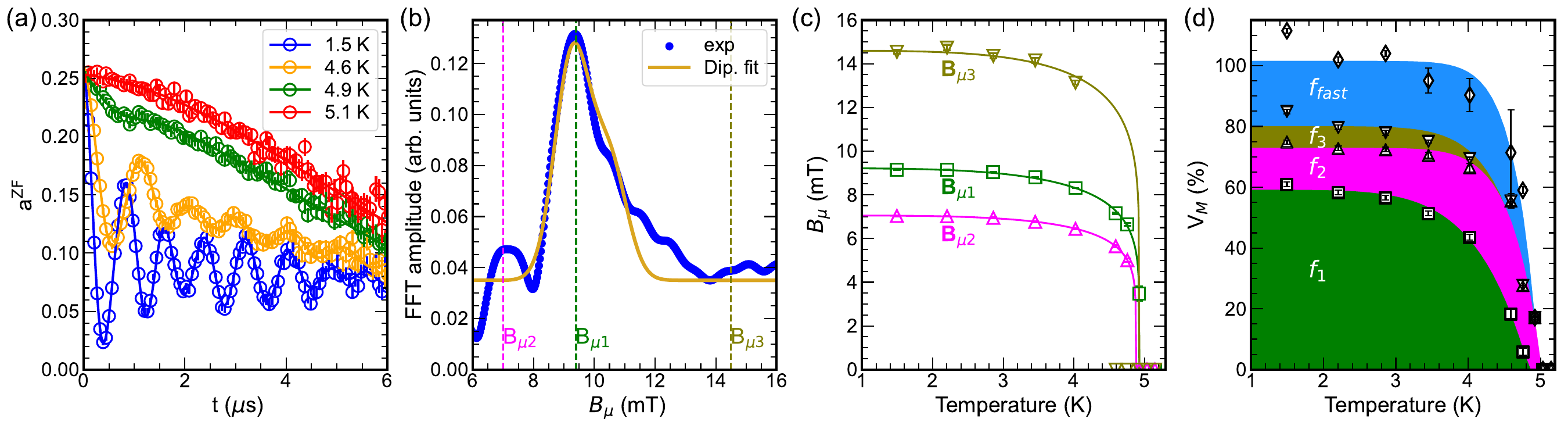}
\caption{(Color online) (a) Muon asymmetry spectra at different temperatures. The solid lines are the best fits to the asymmetry data using Eq.~\ref{formula:fit}. 
(b) The real part of the Fast Fourier Transform (FFT) of the experimental $\mu$SR spectrum at $T\!=\!1.5$~K, with the vertical dashed lines showing the oscillating fields obtained from the best fit of the asymmetry. Solid yellow line is the best fit with the convolution of the computed dipolar fields to a Gaussian distribution with the N\'eel-I AF magnetic structure (See Fig.~\ref{fig:mucalc}(c)). The fit gives a magnetic moment size of $m_{\rm static}\!=\!0.131~\mu_{\rm B}$. (c) The internal fields ($B_{\mu i}$) as a function of temperature. Solid lines are a fit to the power law function described in the text. (d) The magnetic volume fraction $V_{\rm M}$ as a function of temperature, color shaded to illustrate the relative contribution of each term in Eq~\ref{formula:fit}. 
}
\label{fig:musrfit_results}
\end{figure*}

\section{Results}
    \subsection{\label{sec:muon}Muon spin relaxation (\texorpdfstring{\muon}{})}
The measured ZF \muon{} asymmetry spectra, $a^{\mathrm{ZF}}$(t), are shown in Fig.~\ref{fig:musrfit_results} (a) for four different temperatures. %
At temperatures above $\sim$4.9~K, the \muon{} signal shows no oscillations, indicating that the sample is in the paramagnetic phase. Upon lowering the temperature below 4.9~K, the \muon{} signal displays damped coherent oscillations, indicating the emergence of long-range magnetic order due to the presence of a static internal magnetic field. Furthermore, we also observe a fast relaxation at short times, reflecting presence of static magnetic moments. %
 As the temperature is further lowered to 1.5~K, the signature oscillations of the long-range magnetic order become more pronounced and dominate the \muon{} spectra.

All observed ZF-\muon{} spectra are well fit with the following model, commonly used to describe \muon{} asymmetry in antiferromagnets \cite{blundell2022}; 
\begin{equation}
\begin{split}
    a^{\mathrm{ZF}}(t) = &a_0 \biggl[  \sum_{i=1}^{3}{f_ie^{-\lambda_{i} t}\mathrm{cos}(\gamma_{\mu}B_{\mathrm{\mu}, i}^{} t+\phi) }+ \\ &f_{\mathrm{fast}}e^{-\lambda_{\mathrm{fast}} t} + f_l e^{-\lambda_{\mathrm{l}} t}e^{-(\sigma t)^2}  \biggl],
\label{formula:fit}
\end{split}
\end{equation}
where $a_0$ is the muon initial amplitude calibrated at high temperature. In a powder sample we expect the internal local field $\mathbf{B}_\mu$ to have both longitudinal and transverse components with respect to the muon spin $\mathbf{S}_\mu$.
The first term in Eq.~\ref{formula:fit} %
 describes the transverse components with an exponentially damped oscillating signal that decays with a transverse relaxation rate $\lambda$, summed over all muons that thermalize at three symmetrically inequivalent sites. The parameter $f_i$ controls the contribution of each muon stopping  site $i$ to the total asymmetry signal. $\gamma_{\mu}$ (= $2\pi \times 135.5$ MHZ~T$^{-1}$) is the muon gyromagnetic ratio, $B_{\mu}$ (= $2\pi \nu_\mu /\gamma_{\mu} $) is the muon internal magnetic field corresponding to muon precessing with frequency $\nu_\mu$, and $\phi$ is the initial phase.  The second term in Eq.~\ref{formula:fit} accounts for the transverse component for a site with overdamped oscillations, labeled as fast non-oscillating term, with amplitude $a_0 f_{\mathrm{fast}}$ and depolarization rate $\lambda_{\mathrm{fast}}$ ($\sim$ 2 $\mu s^{-1}$). The third term accounts for the longitudinal component with damped relaxation, amplitude $a_0 f_{{l}}$, and spin-lattice relaxation rates $\lambda_{\mathrm{l}}$. Finally,  for  $T \gtrsim T_N$,  static dipolar interactions with nuclear moments with depolarization rate $\sigma\approx 0.2~\mu s^{-1}$ dominate the muon spectra.

In Fig.~\ref{fig:musrfit_results} (b), we show the real part of the Fast Fourier Transform (FFT) of the time-domain \muon ~asymmetry at $T = 1.5$~K.  Three local magnetic fields %
are identified from the best fit of  Eq.\ref{formula:fit} to the asymmetry signal at $T\!=\!1.5$~K. We label these $B_{\mu1}\!=\! 9$~mT, corresponding to the peak with maximum Fourier power, then $B_{\mu2}\! =\! 7$~mT, and  $B_{\mu3}\!=\! 15$~mT. 

The temperature dependence of the three distinct internal fields $B_{\mu i}$  are shown in Fig.~\ref{fig:musrfit_results}(c).  Here the solid lines represent fit to a phenomenological double exponent power-law function; $B_{\mu i}(T)= B_{\mu i}(0) (1 - (\frac{T}{T_N})^\alpha)^\beta$~\cite{PhysRevB.97.224508,LE1993405}. From this fit,  we obtained an estimate of the N\'eel temperature of $T_N\!\approx\!4.9$~K, consistent with heat capacity and neutron measurements~\cite{HINATSU2006155, PhysRevB.103.L121109}. The $\alpha$ exponent accounts for the magnetic excitation at low temperatures while the  $\beta$ exponent reflects the dimensionality of the interactions in the vicinity of the transition.  %
A least squared fit to each internal field component yielded  $\alpha$ = ($3.673 \pm 0.533 $, $4.078 \pm 0.206 $, $4.078 \pm 0.206$) and $\beta$= ($0.160 \pm 0.011$, $0.147 \pm 0.007$, $0.147 \pm 0.007$) for ($B_{\mu1}$, $B_{\mu2}$, $B_{\mu3}$)  respectively.  To obtain an estimate of these exponents we take the average, to find %
 $\alpha = 3.94 \pm 0.20 $ and $\beta = 0.151 \pm 0.005$. 
The $\beta$ value is far less than $\approx$1/3 expected for a three-dimensional magnetic Hamiltonian~\cite{PhysRevB.65.144520}. Its average value is close to that  expected for a two-dimensional XXZ model \cite{Pratt_2007}. However,  the large $\alpha$ value indicates presence of complex magnetic interactions~\cite{PhysRevB.88.134416, PhysRevLett.120.237202}. These findings are in agreement with those from neutrons presented below.

 We obtained the magnetic volume fraction ($V_{\rm M}$) from the asymmetry amplitudes extracted from fits to Eq.~\ref{formula:fit} \cite{PhysRevB.84.195123}. Temperature dependent $V_{\rm M}$ for each internal field component are shown in Fig.~\ref{fig:musrfit_results} (d); the fill color denotes the relative contributions of the muon at each distinct stopping site. The signal $f_1$ corresponds to the muon at site A$_1$ and contributes  59\%  to $V_{\rm M}$ (green shaded area). Sites A$_2$ (signal contribution $f_2$) and A$_3$ (signal contribution $f_3$) contribute 14\% and 7\% respectively, resulting in a total of 80\% contribution to $V_{\rm M}$ from implanted muons that are sensitive to the internal magnetic field. The remaining 20\% is recovered from the amplitude of the fast non-oscillating relaxing signal ($f_{\mathrm{fast}}$), showing that 20\% of the implanted muons do not exhibit coherent precession.

\subsubsection{\label{sec:mucalc}Muon sites and dipolar field analysis}
\begin{figure}[ht!]
\centering
\includegraphics[width = 1.0\linewidth]{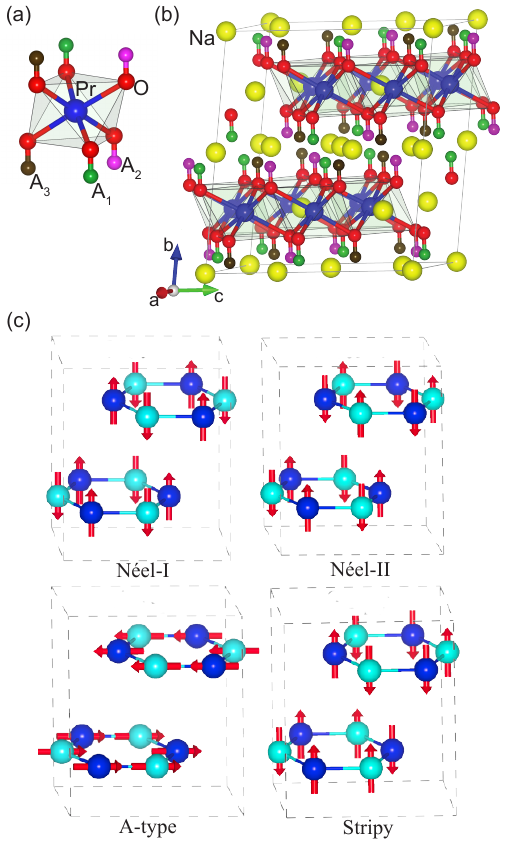}
\caption{ (a) Muon sites A$_1$ (green spheres), A$_2$ (pink) and A$_3$ (brown) bound to the three O sites of the PrO$_6$ octahedra. (b) Muon sites A$_1$ (green spheres), A$_2$ (pink), and A$_3$ (blue) shown in the unit cell of \na{} with lattice parameters $a$ = 5.96 \AA, $b$ = 10.32 \AA, $c$ = 11.73 \AA, $\beta$ = 109.96$^{\circ}$~\cite{ramanathan2021}.  (c) Four proposed AF magnetic structures from analysis of \muon{} data, labelled N\'eel-I, N\'eel-II, A-type, and Stripy. Only the Pr atoms are shown for clarity (with Pr$_1$ dark blue spheres and Pr$_2$ sky blue spheres).}
\label{fig:mucalc}
\end{figure}

To characterize the contributions of each muon site to the oscillatory components of the \muon{} signal in Fig.~\ref{fig:musrfit_results} (a), and constrain the magnetic structure, we proceed to identify the muon implantation site(s) in \na{} using the DFT+$\mu$ approach. 
DFT modeling reveals three symmetrically distinct candidate muon sites, consistent with sites A$_1$, A$_2$, and A$_3$ from the analysis of \muon{} data above. Each of these sites is located at the 8f Wyckoff position, with a distance of $\approx$ 1~\AA{} along the $c$ axis to the three distinct O sites in the unit cell, as shown in Fig.~\ref{fig:mucalc} (a) and (b). These sites are found to be in the direction of the non-magnetic Na layer and farther away from the magnetic Pr$^{4+}$ ions (Fig.~\ref{fig:mucalc} (a)). This is consistent with  observations in typical oxide compounds where  the positive muon is well known to stop near the O sites~\cite{PhysRevB.27.5294,blundell2023}.  

Our DFT calculations predict that the A$_1$ site  has the lowest energy, but the energy differences between A$_1$, A$_2$, and A$_3$ sites are less than 0.2~eV. These findings imply that muons populate the A$_1$, A$_2$, and A$_3$ sites with nearly equal probability.
 
\begin{table}[ht!]
    \centering
    \caption{The best fit magnetic orders labelled N\'eel-I, N\'eel-II, A-type, and Stripy, shown in Fig.~\ref{fig:mucalc} (c); their propagation vector, Group (BNS), their magnetic moment (m) values and the calculated dipolar field ($B_{\rm dip}$) at muon sites A1, A2, and A3 using the corresponding magnetic configuration. $B_{\rm dip}$ values are to be compared with $B_{\mu1}\!=\! 9$~ mT, $B_{\mu2}\! =\! 7$~ mT and $B_{\mu3}\!=\! 15$~mT from experiment.}
    \label{tab:tablemagstr}
    \begin{tabular}{c  c c  c c c c}
    \hline
    \hline
         Label  &  $k_m$  & Group (BNS) & $m_{\rm static}$ ($\mu_\mathrm{B}$)  & $B^{A1}_{\rm dip}$(mT) & $B^{A2}_{\rm dip}$(mT) & $B^{A3}_{\rm dip}$(mT)\\
          \hline
         N\'eel-I  & 000   & C2$'$/c$'$ (15.89) & 0.131  & 11 &  9 &  9 \\
         N\'eel-II   & 000  & C2$'$/c (15.87) & 0.130  & 10 & 10 & 9\\
         A-type   & 000  & C2/c$'$ (15.88) &  0.212&  9& 9 & 10 \\
         Stripy  & 100   & P$_C$2$_1$/c (14.84)$^2$& 0.106 & 9 & 11 & 10\\
\hline
    \end{tabular}
\end{table}

With the knowledge of the muon implantation sites, we determined the magnetic structure and ordered moment size in Na$_2$PrO$_3$ by direct comparison of simulated dipolar field distributions at these site(s) with experimental results. First, by considering the maximal magnetic space groups (MAXMAGN \cite{doi:10.1146/annurev-matsci-070214-021008}), we have identified and explored twenty-eight (28) AF magnetic structures,  within the $k_m$=(000), (110), and (100) propagation vectors that are most likely based on the minimum of the magnetic excitation spectra discussed in Section~\ref{sec:spinwave_neutron}. For each of these magnetic structures, the muon dipolar interactions~\cite{muesr2018} were computed at the 24 muon sites (i.e., 3 muon sites at each of the 8f Wyckoff positions) as a function of the Pr magnetic moment. The experimentally observed FFT power spectrum at 1.5~K (Fig~\ref{fig:musrfit_results} (b)) was then fit to the convolution of the computed dipolar fields and a Gaussian distribution (See Appendix Sec.~\ref{sec:magdip}). The \muon{} magnetic structural analysis identified four possible AF magnetic structures, labeled N\'eel-I, N\'eel-II, A-type, and Stripy in Fig.~\ref{fig:mucalc} (c),  that are consistent with  our experimental findings. We found that the fit to the N\'eel-I AF structure with magnetic moments aligned parallel to the crystal $c-$axis gives the best matching dipolar field distribution shown in Fig.~\ref{fig:musrfit_results}(b). We point out that this magnetic configuration is also in agreement with  the analysis of the neutron data, as we discuss below. Furthermore, as shown in Table~\ref{tab:tablemagstr}, the effective static magnetic moment size determined from the best fit to the real FFT spectrum are notably very small, $m_{\rm static}\!<\!0.22$~$\mu_\mathrm{B}/\mathrm{Pr}$.

\subsection{\label{sec:cf_neutron}Crystal Field Excitations}

\begin{figure*}[t]
\includegraphics[width = 1.0\linewidth]{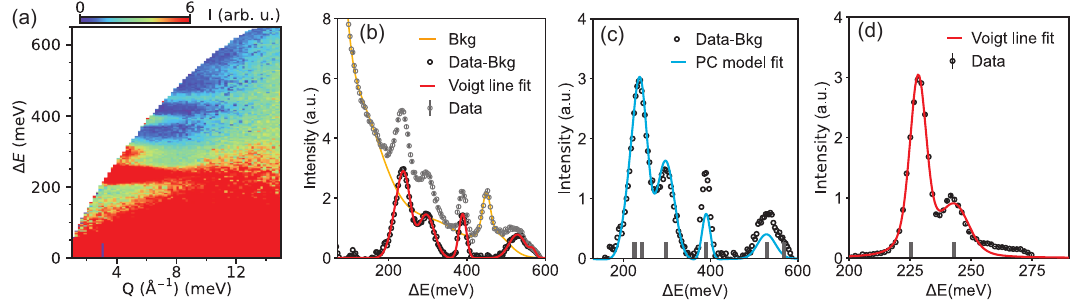}
\caption{ (a) $E_i= 700$~meV inelastic neutron scattering signal in \na{}. (b) Constant momentum transfer cut from $E_i= 700$~meV  data integrated over $|Q|=[6,11]$~\AA$^{-1}$. A background arising from phonons, vibrational excitations and H scattering was subtracted prior to analysis of the crystal field features. (c) Point charge model fit to $E_i = 700$~meV INS spectrum. Vertical bars correspond to the transition levels obtained from the PC model fit.  (d) Constant momentum cut from $E_i=300$~meV data integrated over $|Q|=[6,11]$~\AA$^{-1}$  demonstrating two crystal field levels.}
\label{fig:CEF}
\end{figure*}

In order to elucidate the magnetic degrees of freedom in \na{}, we first present an analysis of crystal field excitations. Fig.~\ref{fig:CEF} (a) shows the low temperature (5~K) inelastic neutron scattering (INS) contour plots with incident energy $E_i= 700$~meV, which clearly reveals at least four inelastic features with intensities that decrease with increasing momentum transfer, consistent with crystal field excitations at $\sim 250$, 300, 400 and 500~meV. The additional local excitation that appears at 452~meV has an intensity that increases with momentum transfer indicating that it is of nuclear origin. We identify this feature with an O-H stretching mode~\cite{LUTZ1982221,PhysRevLett.101.065501}. The low $Q$ magnetic intensity exists on a large high energy background with dispersion that is characteristic of a hydrogen recoil, providing further evidence of the presence of hydrogen contamination that likely originates from a brief exposure of the sample to atmosphere during synthesis. Such O-H stretching modes and hydrogen recoil scattering were also apparent in previous studies of \na{} \cite{PhysRevB.103.L121109}. We note that bulk characterization of our sample as well as neutron diffraction and inelastic scattering presented below are consistent with previous reports, demonstrating that this small amount of hydrogen present has negligible effect on the sample properties. 

To better resolve the intrinsic crystal field excitations from the large background signal, we integrate the neutron data over $|Q|=[6,11]$~\AA$^{-1}$ and fit a background signal using a decaying exponential with an additional Gaussian centered on the 450~meV O$-$H stretching mode as shown in Fig.~\ref{fig:CEF} (b). This background was subtracted to produce the data shown in Fig.~\ref{fig:CEF} (c), while the higher resolution $E_i= 300$~meV data shown in Fig.~\ref{fig:CEF} (d) revealed that the crystal field intensity spectrum below 250~meV is comprised of two modes around $\sim230$ and $\sim240$~meV. No additional crystal field modes were found at higher energies for data collected using neutron incident energies (up to 2.5~eV). To obtain the peak positions, we fit the INS data to a Voigt profile after subtracting a large background (see Fig.~\ref{fig:CEF} (b)). In total, six CEF excitations were revealed at energies of 228, 243, 296, 388, 528 and 568~meV.  The first two modes, at 228 and 243~meV are consistent with prior studies that found a single broad energy mode centered at  233~meV ~\cite{Ramanathan2023, PhysRevB.103.L121109}, but our higher energy resolution clearly shows that this mode is split, while an improved signal to noise unveils four additional crystal field modes that were not previously visible above background.

The six crystal field excitations we observe are consistent with the previously proposed intermediate coupling scheme for \na{} \cite{Ramanathan2023}, confirming the importance of crystal electric field (CEF) interactions in \na{}. However, the higher quality of our data allows us to better constrain microscopic parameters by fitting solely the crystal field level energies and intensities. 

We computed Pr$^{4+}$ single ion crystal electric field (CEF) Hamiltonian ($H_{CEF}$) from a point charge (PC) model within the intermediate coupling regime using the following CEF Hamiltonian~\cite{Scheie-in5044}
\begin{equation}
\begin{split}
H_{CEF} = & B_2^0 O_2^0 + B_2^{\pm 2} O_2^{\pm 2} + B_4^0 O_4^0 + B_4^{\pm 2} O_4^{\pm 2} + B_4^{\pm 4} O_4^{\pm 4} \\
& + B_6^0 O_6^0+ B_6^{\pm 2} O_6^{\pm 2} + B_6^{\pm 4} O_6^{\pm 4}+ B_6^{\pm 6} O_6^{\pm 6},
\end{split}
\end{equation}
where $B_n^{\pm m}$ are the CEF parameters and  $O_n^{\pm m}$ are the Stevens Operators. $H_{CEF}$ includes contributions from the non-zero $-m$ components that are \emph{imaginary} in the Stevens Operators and second order terms. Both are required to account for the low symmetry Pr ion local environment ~\cite{Ramanathan2023}. We also found that  fourth and sixth order terms were required to capture all observed crystal field excitations, especially at higher energy. For our calculations, we considered  Pr$^{4+}$ ion with  orbital angular momentum $L = 3$, spin quantum number $S = 1/2$,  represented by isoelectronic  Ce$^{3+}$ ion.  We obtained reasonable fits of the PC model to the INS data varying the spin orbit coupling strength of between 40 and 60~meV and we report the best fit results for  $\lambda=54$~meV. 

Our model calculations were initialized by generating the PC model to provide a set of starting CEF parameters that were further refined through least-squares fitting to the $E_i=700$~meV INS data. The first column of Table~\ref{tab:bvalues} presents the values of the CEF parameters $B_n^{\pm m}$ from the PC model, while the second column presents the final optimized CEF parameters. A notable feature from the obtained CEF parameters is that significant sixth order terms are required to capture all features of the excitations.

\begin{table}[ht!]
    \centering
    \caption{ $B_n^{\pm m}$ values in meV obtained from the PC model (first column), and fit of the obtained PC model parameters to the INS data at $E_i= 700$~meV.}
    \label{tab:bvalues}
    \begin{ruledtabular}
    \begin{tabular}{c  c  c }
         $B_n^{\pm m}$ &  PC  &  PC fit \\
          \hline
         $B_2^0$       &  -2.1967  &  6.3207\\
         $B_2^{2}$     & -10.4081  & -2.3956\\
         $B_2^{- 2}$   &  2.3018   & 10.1864\\
         $B_4^{0}$     & -0.0800   & -0.0057\\
         $B_4^{2}$     & -0.4789   & -0.2748\\
         $B_4^{- 2}$   & -1.6177   & -1.1675\\
         $B_4^{ 4}$    & -0.7456   & -0.0317\\
         $B_4^{- 4}$   & 0.8509    & 0.1006\\
         $B_6^{0}$     & 0.0080    & -0.0313\\
         $B_6^{ 2}$    & -0.0100   & 0.0890\\
         $B_6^{- 2}$   & -0.0032   & -0.0121\\
         $B_6^{ 4}$    & 0.0625    & -0.0065\\
         $B_6^{-4}$    & 0.0001    &  0.1596\\
         $B_6^{6}$     & 0.0665    & 0.1956\\
         $B_6^{-6}$    &-0.0140    &-0.2250  
    \end{tabular}
    \end{ruledtabular}
\end{table}
   
\begin{table}
\centering
\caption{The ground state eigenvalue (along the first row) and the corresponding coefficients of the eigen-kets (in the $|m_l, m_s\rangle$ basis)  of the fourteen (14) basis sets from the mixing of the $f$ orbitals (down the column), obtained from the point-charge model fit to INS data.} 
\label{tab:eigev}
\begin{ruledtabular}
\begin{tabular}{lcc}
  $\Delta E$ (meV) $\rightarrow$                        & 0.000           & 0.000 \tabularnewline
 \hline 
$|-3,-\frac{1}{2}\rangle$ & (-0.086+0i)     &  0i     \tabularnewline
 $|-3,\frac{1}{2}\rangle$ &  0i             &  (0.143+0i)             \tabularnewline
 $|-2,-\frac{1}{2}\rangle$ & 0i             &  (0.017-0.027j)            \tabularnewline
 $|-2,\frac{1}{2}\rangle$  & (0.155+0.163i) & 0i  \tabularnewline
 $|-1,-\frac{1}{2}\rangle$ & (-0.346-0.478i) & 0i  \tabularnewline
 $|-1,\frac{1}{2}\rangle$  & 0i              & (0.444+0.562i)             \tabularnewline
 $|0,-\frac{1}{2}\rangle$  & 0i              & (-0.17-0.173i)             \tabularnewline
 $|0,\frac{1}{2}\rangle$   & (0.17-0.173i)   & 0i  \tabularnewline
 $|1,-\frac{1}{2}\rangle$  & (-0.443+0.563i) & 0i \tabularnewline
 $|1,\frac{1}{2}\rangle$   &  0i             & (0.345-0.479i)             \tabularnewline
 $|2,-\frac{1}{2}\rangle$  &  0i             & (-0.155+0.163i)             \tabularnewline
 $|2,\frac{1}{2}\rangle$   &  (-0.017-0.027i) & 0i  \tabularnewline
 $|3,-\frac{1}{2}\rangle$  & (-0.143+0i)      & 0i  \tabularnewline
 $|3,\frac{1}{2}\rangle$   & 0i               & (0.086-0i)              \tabularnewline
\end{tabular}
\end{ruledtabular}
\end{table}

Fig.~\ref{fig:CEF} (c) displays a comparison of the  INS intensity with the PC model fit (brown dashed line) that provides excellent agreement with all six observed excitations. The ground state eigenvectors in the $|L, S\rangle$ basis resulting from the PC model fit are reported in Table~\ref{tab:eigev}, while the complete eigenvalues and eigenvectors can be found in the Supplemental Material~\footnote{See Supplemental Material at [URL will be inserted by publisher]. The Supplemental Material includes Refs.~\cite{Scheie-in5044}}.  The low symmetry local environment of Pr$^{4+}$ results in a ground state wavefunction that deviates significantly from the $\Gamma_7$ doublet \cite{PhysRevB.99.241106,PhysRevMaterials.4.104420}. Furthermore, we find a mixed ground state, with significant contribution from the  $j=5/2$ and $j=7/2$ multiplets. The coefficients of all the 14-fold degenerate basis are non-vanishing, as shown in Table~\ref{tab:eigev}.  
The complexity of the ground state wavefunction reflects the low symmetry of the Pr local environment and likely significant hybridization between Pr (4$f$) and O (2$p$) states so that a point charge crystal model provides only an effective description of the magnetism in \na{}. This conclusion is consistent with recent x-ray absorption spectroscopy measurements \cite{Ramanathan2023} and with a reduced local moment as revealed by our \muon{} and neutron scattering measurements presented below. 

The observed $g$-tensor is anisotropic with average transverse components in the honeycomb plane components $g_\pm =  1.05$  and a longitudinal component of  $g_\mathrm{z} = 1.35$, with $g_\mathrm{z}/g_\pm = 1.29$. These illustrate that the CEF effects introduce an easy axis anisotropy that is consistent with a $c-$axis oriented ordered moment.

\subsection{\label{sec:elastic_neutron} Neutron Diffraction}
\begin{figure}[h]
\centering
\includegraphics[width=0.9\columnwidth]{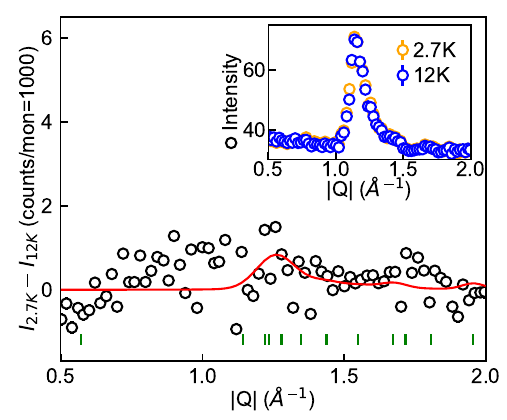}
\caption{Elastic neutron scattering at $T$=2.7~K, well below $T_N$, and 12~K, well above $T_N$ (inset) and the difference between the 2.7~K, and 12 K. The red line is the simulated diffraction pattern for a N\'eel-I order with a moment size of 0.3 $\mu_B$. Expected Bragg peak positions are indicated by the green ticks.}
\label{fig:neut_difrac}
\end{figure}
Both heat capacity \cite{HINATSU2006155} and our \muon{} measurements indicate that \na{} undergoes long-range magnetic ordering below $T_N = 4.9$~K, this ordered state should give rise to a magnetic Bragg reflection visible with neutron diffraction. Fig.~\ref{fig:neut_difrac} shows the elastic neutron signal measured at $T=2.7$~K and $T=12$~K.  To place tight constraints on any ordered magnetic moment, we plot the difference between $T = 2.7$ and 12~K. The difference data do not show any evidence for magnetic Bragg intensity or diffuse scattering above the statistical noise indicating that any three dimensional ordered moment  in \na{} is extremely small, consistent with the \muon{} analysis presented above and previous reports \cite{PhysRevB.103.L121109}. The variance of the difference data places an upper bound on the ordered moment of $m \leq 0.3~\mu_B$ assuming a three-dimensional ordered N\'eel-I magnetic structure with moments aligned along the $c-$axis, consistent with the $\mu$SR and inelastic neutron spectra described below. This upper bound represents a significantly reduced moment compared with expectations from the $J=1/2$ ground state doublet and $g_z$=1.35 from our crystal field analysis, giving  $m_z \approx 0.65~\mu_B$.

Although $\mu$SR clearly reveals a static, ordered, moment below $T_N$ in Na$_2$PrO$_3$, both $\mu$SR and elastic neutron scattering find that this ordered moment is vanishingly small. There are three most likely, and not mutually exclusive, possibilities for such a small ordered moment. The first possibility is the presence of structural disorder that inhibits the formation of three-dimensional long-range order. In particular, stacking faults that are known to be prevalent in Na$_2$PrO$_3$ \cite{ramanathan2021}. The presence of such stacking faults would  increase the upper bound on the ordered moment size determined by neutron diffraction.  The second possibility is significant Pr$-$O covalency that results in a fraction of the moment on the O site.  The third is presence of significant quantum fluctuations that reduce the ordered moment size. Such fluctuations would be expected to arise from a frustrated magnetic Hamiltonian.  Since quantum fluctuations of the ordered moment act to shift magnetic neutron intensity from the elastic ($E=0$) channels to inelastic ones, the possible presence of significant quantum fluctuations is directly testable through analysis of the low energy magnetic excitation spectra as we will discuss below.

\subsection{\label{sec:spinwave_neutron} Magnetic Excitations}
\begin{figure*}[ht]
\centering
\includegraphics[]{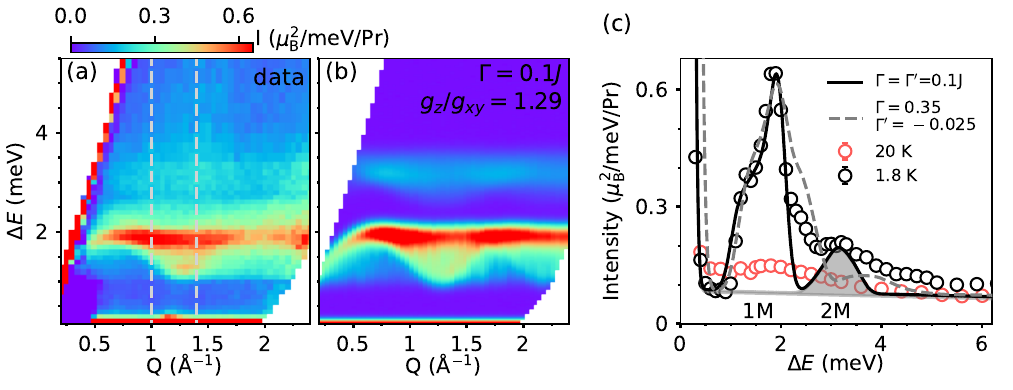}
\caption{(a) Inelastic neutron scattering  data for \na{} at  $T=1.8$~K. (b) Simulated powder averaged non-linear spin wave spectrum for the J-$\Gamma$ model with J=1~meV and $\Gamma = \Gamma^{\prime} = 0.1 J$ corresponding to the XXZ limit. (c) $T = 1.8$ and 20~K constant momentum transfer cut integrated between $1<Q<1.4$ \AA$^{-1}$, as indicated by dashed lines in (a) . Solid line shows the non-linear spin wave model as in (b) with the single (1M) and two-magnon (2M) contributions labelled. Dashed line is the non-linear spin wave model for  $J = 1.24$, $\Gamma = 0.35J$, $\Gamma^{\prime} = -0.025 J$.}
\label{fig:neutron_inelastic_exp}
\end{figure*}

The measured powder averaged low energy ($\delta E<10$~meV) inelastic neutron scattering for \na{} is shown in Fig.~\ref{fig:neutron_inelastic_exp} (a). There are two visible branches of magnetic excitations, centered at 1.5 and 3~meV. The lower energy branch exhibits a clear dispersion, with a 1~meV gap and 1.5~meV bandwidth consistent with previous reports \cite{PhysRevB.103.L121109}. Excitations disperse from a minimum energy at $Q=1.25$~\AA$^{-1}$ corresponding to the $(110)$ Bragg position as expected for spin waves in the N\'eel-I ordered state. The higher energy branch of magnetic excitations is centered around  3.2~meV and extends into a continuum up to 5~meV, most clearly visible in the constant momentum transfer cut of Fig.~\ref{fig:neutron_inelastic_exp} (c). Given that this high energy branch intensity is maximum at approximately twice the energy of the lower energy zone boundary --  where the density of single magnon states is maximized -- and the intensity of the higher energy branch is significantly reduced with respect to the lower energy one, we associate the higher energy excitations with a multi-magnon continuum. Such an assignment is supported by non-linear spin wave modeling discussed below. Both branches of magnetic excitations follow the same temperature dependence, collapsing into a broad energy continuum of the  paramagnetic response above $T_N$.

To model the magnetic excitations in the N\'eel state, we consider the generic nearest-neighbour model for Kramer's pseudo-spins on the honeycomb lattice. This includes four symmetry allowed exchange interactions~\cite{rau2014,katukuri2014kitaev}, 
\begin{align*}
\sum_{\langle{ij}\rangle_{\mu}}&\left[
J\vec{S}_i \cdot \vec{S}_j + K S^{\mu}_i S^{\mu}_j 
+ \Gamma \left(S^{\nu}_i S^{\rho}_j +S^{\rho}_i S^{\nu}_j\right)
\right. \\&\left. 
+ \Gamma' \left(S^{\mu}_i S^{\rho}_j +S^{\mu}_i S^{\nu}_j+
+ S^{\rho}_i S^{\mu}_j +S^{\nu}_i S^{\mu}_j\right)
\right],    
\end{align*}
where $\langle{ij}\rangle_\mu$ is a $\mu$-type bonds and $\mu\nu\rho$ are a permutation of the octahedral axes $xyz$. This includes the Heisenberg exchange $J$, Kitaev exchange $K$ and two symmetric off-diagonal exchanges $\Gamma$ and $\Gamma'$. We will assume the strictly crystallographically inequivalent nearest neighbor bonds have the same exchange interactions, effectively granting the model three-fold rotation symmetry about the $c$-axis. We do not include $c-$axis exchange interactions as these are expected to be weak and minimally influence the powder averaged dynamical structure factor.

We have assumed the $J=1/2$ pseudo-spins in \na{} are defined with respect to the local cubic axes. Alternatively, a trigonal basis that quantizes the spins along the out-of-plane and high-symmetry in-plane directions can be used and yields~\cite{rau2018yb, maksimov2020}
\begin{align*}
\sum_{\langle{ij}\rangle} &\left[
J_1 \left({S}_i^x {S}_j^x+ {S}_i^y {S}_j^y + 
\Delta {S}_i^z {S}_j^z\right)
+J_{\pm\pm}\left( \gamma_{ij} {S}^+_i {S}^+_j +{\rm h.c.}\right)\right. \\
&\left.-J_{z\pm}\left( \gamma^*_{ij}\left[ {S}^+_i {S}^z_j+ {S}^z_i {S}^+_j\right] +{\rm h.c.}\right) \right],
\end{align*}
where $\gamma_{ij}$ are bond dependent phase factors ($\gamma_x = 1$, $\gamma_y = \omega$, and $\gamma_z = \omega^*$ where $\omega = e^{2\pi i/3}$).

Classically, a N\'eel state can be stabilized  on the honeycomb lattice by a dominant anti-ferromagnetic exchange $J>0$, leaving the staggered moment direction arbitrary. Including a Kitaev interaction in addition to the Heisenberg interaction stabilizes a staggered moment aligned along one of the cubic axes through an order-by-quantum-disorder mechanism. Finite symmetric off-diagonal exchanges will select a direction even classically, with $\Gamma + 2\Gamma' > 0$ favoring out of the honeycomb plane, corresponding to $\Delta>1$ in the trigonal basis, and $\Gamma + 2\Gamma'<0$, corresponding to $\Delta<1$, favoring an in-plane staggered moment -- the in-plane direction is unfixed classically, but will be selected via order-by-disorder~\cite{rau2018}.  We assume the staggered moment orientation is selected by the symmetric off-diagonal exchanges given the small selection energy of the order-by-disorder mechanism and fix $\Gamma + 2\Gamma'>0$ to yield a staggered moment oriented out of the honeycomb plane, consistent with expectations from $\mu$SR.

The magnetic excitations in the N\'eel phase can be calculated semi-classically using spin-wave theory. We carried out calculations of the powder-averaged dynamical structure factor for several sets of exchange constants $(J,K,\Gamma,\Gamma')$ including the XXZ limits (only $J_1$, $\Delta$ non-zero) and the $J-\Gamma-\Gamma'$ model (setting $K=0$). We note that the XXZ limit is equivalent to $K=0$ and $\Gamma=\Gamma'$ with $J_1 = J-\Gamma$ and $\Delta J_1 = J+2\Gamma$. Given the small ordered moment we have included corrections to linear spin wave theory up to order $O(1/S^2)$ to ensure we can capture any quantum fluctuations. Details of the formalism for these non-linear spin-wave theory calculations can be found (e.g.) in Refs.~[\onlinecite{rau2018, mcclarty2018, rau2019}].

First, we note that large gap observed experimentally can be captured even in linear spin-wave theory with three-fold symmetry once $\Gamma$ and $\Gamma'$ interactions are included (or equivalently, once $\Delta \neq 1$). Quantitatively, the size of the gap, and the bandwidth of excitations, can be accounted for by several different sets of exchange parameters. The key features of the spectrum are largely determined by the value of $\Gamma + 2\Gamma'$ (where $J$ is fixed to set the overall energy scale). Tuning the precise values of $\Gamma$ or $\Gamma'$ separately only introduces subtle changes to the powder-averaged intensity that our data cannot constrain. Based on these gross features, we confine our calculations to the value $\Gamma + 2\Gamma' = 0.3J$, fixing the scale of the gap relative to the bandwidth, and vary the relative contributions of $\Gamma$, $\Gamma'$. Small differences can be observed in the ``flatness'' of the top of the excitation band and in the distribution of intensities near maxima at $|\vec{Q}| \sim 1$ \AA{} and $1.75$~\AA{} depending on the precise values used. 

To model \na{} we considered several cases: $\Gamma/J = 0.0, 0.1, 0.2, 0.3$ and $0.35$ holding $\Gamma'/J = (0.3-\Gamma/J)/2$ for each to fix $\Gamma +2\Gamma'$. This includes the XXZ limit where $\Gamma=\Gamma'=0.1J$ corresponding to $\Delta = 1.33$ in the trigonal basis, close to the value considered in Ref.~[\onlinecite{PhysRevB.103.L121109}]. A comparison of this model to the experimental data is shown in Fig.~\ref{fig:neutron_inelastic_exp} (b) and (c). The theoretical result incorporates up to the $O(1/S^2)$ contributions of the transverse, longitudinal and transverse-longitudinal parts of the structure factor (which include the leading contribution from the two-magnon continuum)~\cite{zhito2013, mourigal2013}. The overall intensity scale is left arbitrary and a $g$-factor anisotropy of $g_z/g_\pm \approx 1.29$ was used. Due to the limited range of $|\vec{Q}|$, the neutron form factor for Pr\textsuperscript{4+} was not included.

Overall the $J-\Gamma$ model with J=1~meV, $\Gamma/J=-0.1$, and  $\Gamma'=\Gamma$  provides an excellent description of the data given the simplicity of the model.  We found no improvement moving away from the XXZ limit or including an additional Kitaev exchange.  Including the symmetry inequivalent nearest neighbor exchange and out-of-plane exchange interactions will likely lead to an improved description of the data, but our powder averaged data set is not sufficient to constrain such a large parameter space and the minimal $J-\Gamma$ model presented here captures the essential physics. Single crystal measurements are required for any reliable further refinement of additional parameters in the magnetic Hamiltonian, to capture the magnetic intensity around $\Delta E=2.5$~meV and above 4~meV.

A key feature of the magnetic excitations is the pronounced continuum with a maximum intensity visible near $\delta E = 3$~meV. Given the powder-averaging, it is tempting to attribute this to an additional band arising due to Kitaev or additional anisotropic interactions. A simpler explanation is that it represents the contribution from the two-magnon continuum associated with the lower energy single magnon band. Kinematically, this is sensible, since the continuum starts above twice the gap, but its unusually high intensity merits further discussion. 

Since the two-magnon intensity is determined by components of the spin along the direction of the ordered moment, any reduction of the ordered moment that results from quantum fluctuations must be accounted for by an enhanced two magnon signal. For the XXZ model most relevant to \na{} we expect only a modest 20\% ordered moment reduction including corrections up to second order, as found in Ref.~\cite{Zheng1991}. Indeed, such a modest moment reduction is anticipated from the data directly, as the large  measured spin wave gap relative to the bandwidth acts to energetically ``freeze out'' moment fluctuations at low temperatures.   We conclude that rather than enhanced quantum fluctuations, the strong two-magnon intensity in \na{} is a result of $g$-factor anisotropy in \na{}.

While the one-magnon intensity is determined by the spin components transverse to the ordered moment, and thus are $\propto g_{\pm}^2$, the two-magnon intensity is sensitive to the longitudinal component and carries a factor of $g_z^2$. The measured $g$-factor ratio $g_z/g_\pm \approx 1.29$ thus provides an enhancement of the two-magnon intensity of $(g_z/g_\pm)^2 \approx 1.66$ relative to the one magnon intensity.  Comparison of the non-linear spin wave dynamic structure factor with our data in Fig.~\ref{fig:neutron_inelastic_exp} (c) shows excellent agreement of the relative two magnon intensity with the measured signal at 3~meV.   

Despite the apparently small static moment and absence of any magnetic Bragg signal, the inelastic spectra displays well-formed magnon excitations that can be described to a high fidelity with non-linear spin wave theory predicting a modest, $\sim$20\% ordered moment reduction from quantum fluctuations. In the absence of significant quantum fluctuations, the unobservable magnetic Bragg intensity in the powder averaged diffraction data  could be accounted for by stacking faults that disrupt three-dimensional magnetic order, allowing long-range order to form in two-dimensional honeycomb planes but only short range order between the planes. The resulting magnetic correlations form  ``rods'' of elastic scattering extending along the $c^{*}$ direction. When powder averaged, the rods result in a diffuse signal that is not visible above background. Since the magnetic interactions along the $c-$axis are weak, the magnetic excitations do not disperse along this direction and the powder averaged inelastic neutron intensity is not significantly influenced by stacking faults. However, such a scenario cannot account for the small static moment determined by \muon{}.

The inelastic neutron scattering signal provides an additional, independent check, of the magnetic moment size in \na{} through the total moment sum rule. Integrating the total measured inelastic magnetic neutron intensity over the region $0.7\!<\!\Delta E\!<6$~meV and $0.8\!<\!Q\!<\!1.94$~\AA$^{-1}$, we obtain an approximate total fluctuating moment of $\delta m^2 \!= 3/2  \mu_{\rm B} ^2\!\int\!\int\!Q^2 I(Q,E) {\rm d}Q{\rm d}E/\!\int\!  Q^2{\rm d}Q\!=\!0.57(22)$~$\mu_{\rm B}^2$/Pr. This value accounts for a large systematic error arising from background determination and is reduced from the value of $\delta m^2_{\rm local} = g_{\rm avg}^2J\left(J+1\right) - \delta_{\rm z}^2g_{\rm z}^2J_z^2 \!=\!0.7$~$\mu_{\rm B}^2$/Pr expected for a local $J=1/2$ moment and $g$-factors as determined by crystal field analysis. The factor $\delta_{\rm z}=0.8$ accounts for a static moment reduction from quantum fluctuations.  Such a reduced fluctuating moment demonstrates the absence of significant quantum fluctuations and is  consistent with $m_{\rm static} \leq 0.2$~$\mu_{\rm B}$ determined by \muon{}. Overall, \muon{}, neutron crystal field measurements, neutron diffraction, and low energy inelastic scattering reveal that \na{} is a well ordered N\'eel antiferromagnet, exhibiting minimal quantum fluctuations, but there is a large moment reduction from crystal field effects that give rise to a small, anisotropic, $g$-factor and covalency that reduces the moment from the single ion limit.

\section{\label{sec:discussion}Summary and Conclusions}
Our combined neutron and \muon{} measurements have demonstrated that \na{} is  N\'eel antiferromagnet  and well described by a two dimensional $J-\Gamma$ or equivalently $XXZ$ model Hamiltonian. Although no magnetic Bragg reflection corresponding to long-range order has been identified, \muon{} measurements reveal clear oscillations characteristic of long-range order. DFT modeling enabled us to identify the muon stopping sites and compare measured internal field distributions against possible magnetic structures. We find the \muon{} internal field distribution is most consistent with  N\`eel order with a small ($\leq 0.22$~$\mu_{\rm B}$) static moment. Non-linear spin wave modeling of the observed collective magnetic excitations capture the complete spectra. An intense two-magnon band is accounted for through $g$-factor anisotropy that acts to enhance the neutron intensity of the longitudinal two-magnon excitations relative to the transverse single magnons, by a factor of $(g_z/g_{\pm})^2 \sim 1.66$. We find minimal reduction of the ordered moment from frustration or quantum fluctuations as confired through an  analysis of the total magnetic spectral weight from inelastic neutron scattering. Overall our work emphasizes how local atomic physics can generate small static magnetic moments and  intense multi-magnon continuum observed by inelastic neutron scattering experiments, even in the absence of appreciable quantum fluctuations. Furthermore, our work demonstrates the importance of \muon{} as a technique for investigating frustrated magnets with small moments and the clear synergies between \muon{} and INS.

\section{acknowledgements}
K.W.P and Q.W. were supported by the U.S.  Department of Energy, Office of Basic Energy Sciences, under Grant No. DE-SC0021223.  VFM was supported by  the National Science Foundation under grant No.  DMR-1905532. JGR was supported by the Natural Sciences and Engineering Research Council of Canada (NSERC) (Funding Reference No. RGPIN-2020-04970). Access to MACS was provided by the Center for High Resolution Neutron Scattering, a partnership between the National Institute of  Standards and Technology and the National Science Foundation under Agreement  No. DMR-1508249.  A portion of this research used resources at the Spallation Neutron Source, a DOE Office of Science User Facility operated by the Oak Ridge National Laboratory. IJO and PB acknowledge financial support from PNRR MUR project ECS-00000033-ECOSISTER and also acknowledge computing resources provided
by the STFC scientific computing department’s SCARF cluster and CINECA award  under the ISCRA (Project ID IsCa4) initiative.

\appendix
\appendix
\setcounter{equation}{0}
\renewcommand{\theequation}{A\arabic{equation}}

\section{\label{sec:magdip}Magnetic structure determination and \muon{} dipolar field simulations}
The search for the magnetic structure in \na{} was initialized by considering all possible magnetic configurations for $k_m=(0, 0, 0)$, (1, 1, 0), and (1, 0, 0) propagation vectors and its crystal structure.  In Table~\ref{tab:table3} we have listed the maximal magnetic space groups and magnetic configurations that we have obtained utilizing the MAXMAGN \cite{doi:10.1146/annurev-matsci-070214-021008}.

\begin{table*}
 \begin{threeparttable}
\centering
\caption[magnetic _symmetry]{\label{tab:table3} Magnetic structures from MAXMAGN - Bilbao Crystallographic Server \cite{doi:10.1146/annurev-matsci-070214-021008}: Maximal magnetic space groups for a given propagation vector $k_m$, with two magnetic Pr atoms, Pr1 and Pr2.}
\begin{ruledtabular}
 \begin{tabular}{ l@{\hspace{0.6em}} p{2.5cm}  p{10.0cm} }
  $k_m$  & Group (BNS)  & Magnetic Structure~\tnotex{tn1} \\  
       \colrule
 $000$                     & C2$'$/c$'$ (15.89)   & Pr$_1$:  (m$_x$, 0, m$_z$), (m$_x$, 0, m$_z$ ), (m$_x$, 0, m$_z$), (m$_x$, 0, m$_z$)     \\
                                 &                                  & Pr$_2$:  (m$_x$, 0, m$_z$), (m$_x$, 0, m$_z$ ), (m$_x$, 0, m$_z$), (m$_x$, 0, m$_z$)     \\
                                  & C2/c$'$ (15.88)        & Pr$_1$:  (0, m$_y$, 0), (0, $-$m$_y$, 0), (0, m$_y$, 0), (0,$-$m$_y$, 0)  \\
                                  &                                  & Pr$_2$:   (0, m$_y$, 0), (0, $-$m$_y$, 0), (0, m$_y$, 0), (0, $-$m$_y$, 0)  \\
                                   & C2$'$/c (15.87)        & Pr$_1$:   (m$_x$, 0, m$_z$), ($-$m$_x$, 0, $-$m$_z$), (m$_x$, 0, m$_z$), ($-$m$_x$, 0,$-$m$_z$) \\
                                   &                                  & Pr$_2$:  (m$_x$, 0, m$_z$), ($-$m$_x$, 0, $-$m$_z$), (m$_x$, 0, m$_z$), ($-$m$_x$, 0,$-$m$_z$) \\
                                   & C2/c (15.85)             & Pr$_1$:   (0, m$_y$, 0), (0, m$_y$, 0), (0, m$_y$, 0), (0, m$_y$, 0) \\
                                   &                                  & Pr$_2$:  (0, m$_y$, 0),  (0, m$_y$, 0), (0, m$_y$, 0),  (0, m$_y$, 0) \\
 \hline                                             
$100$                         & P$_C$2$_1$/c (14.84)     & Pr$_1$:  (m$_x$, 0, m$_z$), (m$_x$, 0, m$_z$ ), ($-$m$_x$, 0, $-$m$_z$), ($-$m$_x$, 0, $-$m$_z$)\\
                                   &                                          & Pr$_2$:   (m$_x$, 0, m$_z$), (m$_x$, 0, m$_z$), ($-$m$_x$, 0, $-$m$_z$), ($-$m$_x$, 0, $-$m$_z$)\\   
                                   & P$_C$2$_1$/c (14.84)     & Pr$_1$:   (m$_x$, 0, m$_z$), ($-$m$_x$, 0, $-$m$_z$) , ($-$m$_x$, 0, $-$m$_z$), (m$_x$, 0, m$_z$) \\
                                   &                                          & Pr$_2$:   (m$_x$, 0, m$_z$), ($-$m$_x$, 0, $-$m$_z$), ($-$m$_x$, 0, $-$m$_z$), (m$_x$, 0, m$_z$) \\
                                   & P$_C$2/c (13.74)             & Pr$_1$:    (0, m$_y$, 0), (0, m$_y$, 0), (0, $-$m$_y$, 0), (0, $-$m$_y$, 0) \\
                                   &                                          & Pr$_2$:   (0, m$_y$, 0), (0, m$_y$, 0), (0, $-$m$_y$, 0), (0, $-$m$_y$, 0)  \\
                                   & P$_C$2/c (13.74)             & Pr$_1$:     (0, m$_y$, 0), (0, $-$m$_y$, 0), (0, $-$m$_y$, 0), (0, m$_y$, 0)  \\
                                   &                                          & Pr$_2$:   (0, m$_y$, 0), (0, $-$m$_y$, 0), (0, $-$m$_y$, 0), (0, m$_y$, 0)                                                                    
    \end{tabular}
\end{ruledtabular}
 \begin{tablenotes}
			\item[a]  \label{tn1} The atomic positions in fractional coordinate units \cite{ramanathan2021} and the Wyckoff positions to the above magnetic order are;\\
			Pr1:  0.00000, 0.16590, 0.25000 - (0, y, 1/4), (0, $-$y, 3/4), (1/2, y+1/2, 1/4), (1/2, $-$y+1/2, 3/4) \\
			 Pr2:  0.00000, 0.50100, 0.25000 - (0, y, 1/4), (0, $-$y, 3/4), (1/2, y+1/2, 1/4), (1/2, $-$y+1/2, 3/4)            
		\end{tablenotes}  
  \end{threeparttable}
  	\
\end{table*}

Starting from these maximal magnetic space groups in Table~\ref{tab:table3}, we have considered only antiferromagnetic structures with the following conditions of the magnetic moments; (i) m$_x$ = m$_z$ = $\sqrt{2}/2 |m|$, (ii) m$_x$ = $|m|$ and m$_z$ = 0; (iii) m$_x$ = 0 and m$_z$ = $|m|$, (iv) m$_y$ = $|m|$ and (v) taking into account  two distinct Pr sites, such that for cases i to iv above, we have considered that the moments on Pr$_1$ is antiparallel to those on Pr$_2$.  Implementing these conditions, we obtained a total of 28 possible magnetic structures.  Due to the magnetic symmetry of the compound, the magnetic structures obtained for the propagation vector $k_m$ = (0,0,0) are the same for $k_m$ = (1,1,0).

\begin{figure}[!h]
\includegraphics[width = 0.5\textwidth]{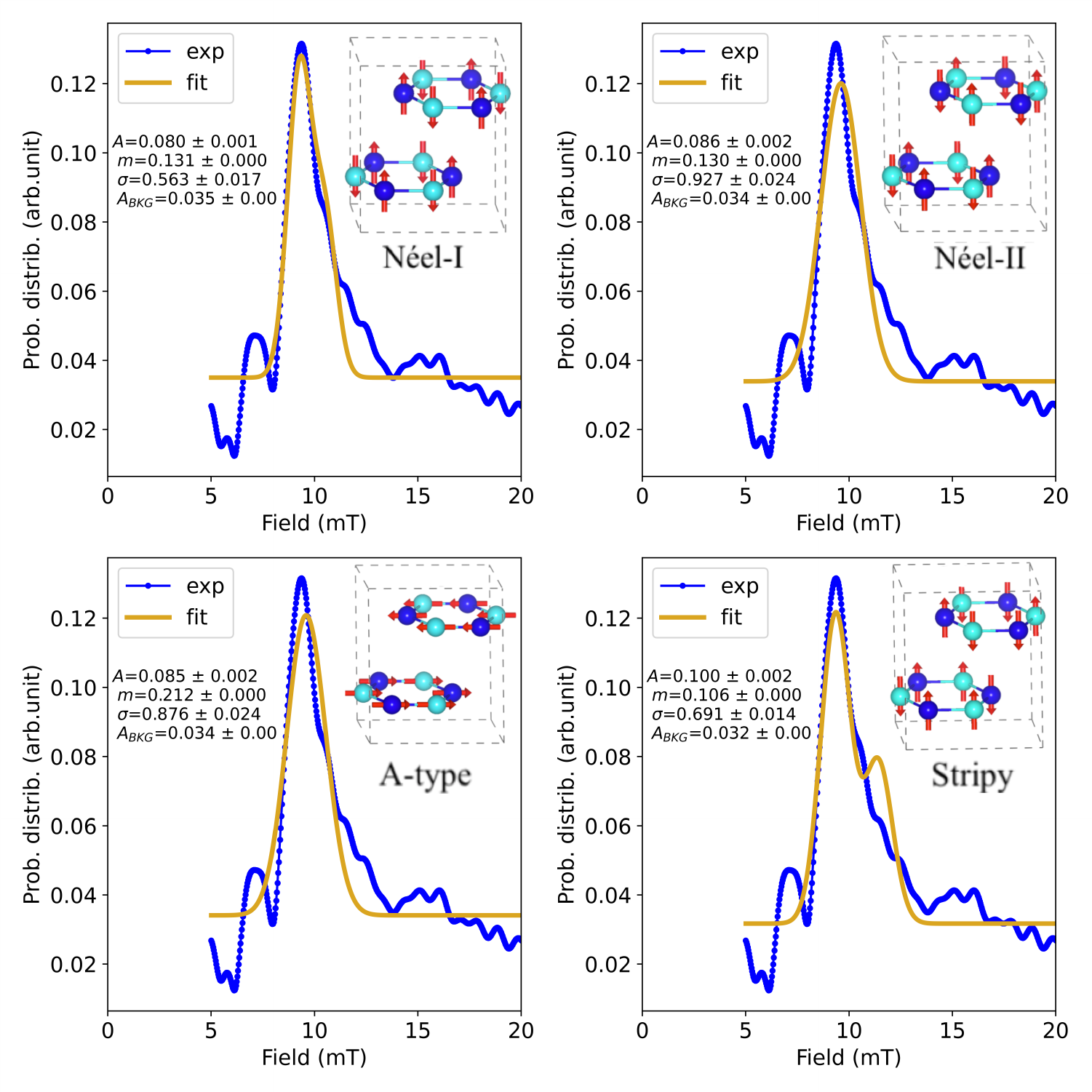}
\centering
\caption[Untc ]{\label{fig:magsStructures} Comparison between experimental and calculated dipolar field distribution for four candidate magnetic structures. Fit parameters from Eqs.~\ref{dipeq1}-\ref{dipeq3} are shown in the inset of each panel.}
\end{figure}
In order to obtain  the magnetic structure and estimate the magnetic moment of Na$_2$PrO$_3$ using the muon spin spectroscopy measurements, we approximated and fit the Gaussian convolution of the muon dipolar fields  calculated over these 28 magnetic structures to the experimentally observed Fourier power spectrum at 1.5~K. The simulation of the dipolar contribution to the muon internal field also requires the knowledge of the muon site~\cite{muesr2018}. As described in the main text, the muon was found to thermalize at three symmetrically inequivalent sites,  that we have labelled A$_1$, A$_2$, and A$_3$ (See Main text Sec.~\ref{sec:mucalc}). Each of these sites has a multiplicity of 8 (8f Wyckoff position), hence we have considered dipolar fields calculated on 24 muon positions.

For each magnetic structure, the dipolar fields $B$~\cite{muesr2018} are calculated for all 24 muon sites with index i,  in the pristine \na{} structure as a function of the Pr moments, $m_{Pr}$, written as

\begin{equation}
p(B, m_{Pr} ) = \sum_{i=1}^{24} \delta(B-m_{Pr} B_i).
\label{dipeq1}
\end{equation}

Such that we can approximate the FFT distribution of the ZF-\muon{} experimental spectra to a convolution with a Gaussian broadening $g$ as

\begin{equation}
\widetilde{p}(B, m_{Pr}, \sigma) = (p \ast g) (B) := \int_{-\infty} ^\infty p(\tau, m) g(B-\tau, \sigma) d\tau.
\label{dipeq2}
 \end{equation}
The fitting function $P$ becomes

\begin{equation}
P(B; m, \sigma, A, A_{BKG}) = A\widetilde{p}(B, m, \sigma) + A_{BKG},
\label{dipeq3}
\end{equation}

where $\sigma$, $A$, $A_{BKG}$ are the width , amplitude and the background of the distribution respectively. Statistically acceptable fits were obtained only for four magnetic configurations shown in Fig.~\ref{fig:magsStructures}. These magnetic structures consist of the  N\'eel-I, N\'eel-II, A-type, and Stripy antiferromagnetic configurations.

\end{document}